# HEAT TRANSFER THROUGH NEAR-FIELD INTERACTIONS IN NANOFLUIDS

Philippe Ben-Abdallah

Laboratoire de Thermocinétique, CNRS UMR 6607, Ecole Polytechnique de l'Université de Nantes, 44 306 Nantes cedex 03, France.

pba@univ-nantes.fr

**Abstract**

Using the Landauer-Buttiker theory we calculate the thermal conductance associated to plasmons modes in one dimensional arrays of nanoparticles closely spaced in a host fluid. Our numerical simulations show that the near-field interactions between particles have a negligible effect on the thermal conductivity of nanoparticles colloidal solutions (nanofluids).

Recent experiments have revealed [1-4] that many colloidal solutions, also called nanofluids, built by dispersing nanometer-size particles in a liquid display an anomalously high thermal conductivity that the classical theory of effective media [5] fails to explain. Several scenarios have been proposed to explain these unexpected results. Hydrodynamics effects due to Brownian motion of particles [6] have been first suggested to explain this enhancement but recently Evans et al. [7] have shown that the nanoscale fluid flow resulting from this motion have only a minor contribution in the thermal conductivity increase of nanofluids. In parallel to this work, a numerical investigation of nanofluids based on molecular dynamic simulations [8] have enabled to demonstrate that only collective effects could explain the heat transfer enhancement in these media.

In this Letter we examine one of such mechanisms. Our work is based on the following observations : (i) Most of anomalous increases in the thermal conductivity has been observed with nanofluids made with metallic or polar nanoparticles that support plasmons or phonon-polariton modes which have resonance frequencies in the visible or near infrared range. (ii) At weak volume fraction ($\phi \sim 1\%$) the mean distance $d_a = [(\frac{\pi}{6\phi})^{1/3} - 1]d$ between two neighborhoods particles in suspension is of the order of their diameter d. Therefore the surface plasmons or phonon-polariton supported by these particles collectively interact in the host fluid. In recent theoretical developments [9,10] some authors have predicted strong near-field coupling between two nanoobjects or between a bulk material and a nanoobject closely spaced by vacuum suggesting so the possibility of a strong enhancement of heat exchanges from this way. Here we estimate the order of magnitude of heat collectively transported by near-field interactions in a host fluid. We will limit ourselves to linear chains of nanoparticles at rest in a dielectric medium. The thermal conductance associated to the interaction of plasmons modes will be calculated using the Landauer-Buttiker theory of transport [11].

To start let us consider a chain of N spherical nanoparticles regularly distributed in a host material along a linear chain as shown in (Fig.1). To determine the thermal conductance of this chain we connect its extremities to two massive materials which are maintained at temperatures *T* and $T+\delta T$, the temperature difference $\delta T$ being small compared to the mean temperature $\frac{2T+\delta T}{2}$. Also we assume that both reservoirs support surface modes which are able to perfectly couple with the chain [9]. Energy exchanged between both reservoirs via the collective excitations of electrons (plasmons) can be calculated using the Landauer-Buttiker theory. Hence, introducing the right and left moving plasmons energy flux

$$\varphi^\pm = \frac{1}{2\pi}\sum_m \int_0^\infty v_{gm}(k)\hbar\omega_m(k)f_B^\pm[\omega_m(k)]dk, \qquad (1)$$

where $k$ stand for the wave vector, $\omega_m(k)$ the dispersion relation of m th mode, $v_{gm} = d\omega_m/dk$ the group velocity of this mode and $f_B(\omega) = [\exp(\beta\hbar\omega)-1]^{-1}$ the distribution function of plasmons (bosons), the thermal conductance of a nanoparticle chain associated to plasmons is given by

$$G = \lim_{\delta T \to 0} \frac{\varphi^+(T+\delta T) - \varphi^-(T)}{\delta T}. \qquad (2)$$

Finally taking into account the plasmons damping along the chain and transforming the integrals in (1) to integrals over frequencies this conductance writes

$$G = \frac{\hbar^2}{2\pi k_B T^2} \sum_m \int_0^\infty \Im_m(\omega)\omega^2 \frac{e^{\beta\hbar\omega}}{(e^{\beta\hbar\omega}-1)^2} d\omega, \qquad (3)$$

where $\Im_m(\omega)$ represents the transmission probability through the chain of m th plasmon mode at the frequency $\omega$. The discrete sum is taken over all polarization states including one longitudinal, non-degenerated, and two transversals, two times degenerated, branches.

Treating the particles chain as a linear distribution of point dipoles involving only the Förster electric field [12] (proportional to $d_a^{-3}$) which is dominant in the quasistatic limit ($d_a \ll \lambda_{plasmon}$), the equation of motion of dipolar moments $p_{m,i}$ writes [13]

$$\ddot{p}_{m,i} = -\omega_0^2 p_{m,i} - \Gamma_I \dot{p}_{m,i} + \frac{\Gamma_R}{\omega_0^2}\dddot{p}_{m,i} - \gamma_m \omega_1^2(p_{m,i-1} + p_{m,i+1}), \qquad (4)$$

where the subscript labels the particles, $\omega_0$ represents the eigenfrequency of oscillating dipoles, $\gamma_m$ is a polarization dependent constant for which $\gamma_T = 1$ for transversal modes and $\gamma_L = -2$ for longitudinal modes. Here $\Gamma_I$ and $\Gamma_R$ represent the electronic relaxation frequencies due to interactions with phonons and electrons and due to radiation into the far field respectively. For metal particles with a diameter small in front of the plasmon wavelength, the dissipation is mainly non-radiative [14] so that $\Gamma_R$ can be neglected in Eq. (4). The dielectric constant of nanoparticles is described by a Drude model (i.e. $\varepsilon(\omega) = 1 - \frac{\omega_p^2}{\omega(\omega + i\gamma)}$, $\gamma = 1/\tau > 0$ being directly related to $\tau$ the average time between two subsequent electron collisions) so that the relaxation frequency is very well described [15] by the following expression

$$\Gamma_I(\omega) = \frac{2\,\mathrm{Im}[\varepsilon(\omega)]}{\left[(\frac{\partial \mathrm{Re}[\varepsilon(\omega)]}{\partial \omega})^2 + (\frac{\partial \mathrm{Im}[\varepsilon(\omega)]}{\partial \omega})^2\right]^{1/2}} \approx \gamma \qquad (5)$$

As for the last term in Eq. (2) it represents the interaction of dipole with its neighbors, $\omega_1^2$ being the coupling strength. The solution of the linear equation (2) is on the damped oscillating form

$$p_{m,i} = \mathrm{P}_{m,0} \exp[-\alpha_m m d_a + i(\omega t \pm k m d_a)] \qquad (5)$$

where $\alpha_m$ is the linear attenuation of m th plasmon mode along the chain. Inserting the ansatz (5) into Eqs. (4) and separating the real and imaginary part of result we find

$$\omega^2 = \omega_0^2 + 2\gamma_m \omega_1^2 \cos(k d_a) \cosh(\alpha_m d_a), \qquad (6\text{-a})$$

$$0 = \omega \Gamma_I \mp 2\gamma_m \omega_1^2 \sin(k d_a) \sinh(\alpha_m d_a). \qquad (6\text{-b})$$

Solving these equations with respect to the damping factor $\alpha_m$ and to the wave number $k$ we obtain after a straightforward calculation

$$\alpha_m = \frac{1}{d_a} \mathrm{ArgSinh}[\frac{\omega \Gamma_I}{2|\gamma_m|\omega_1 \sin(k d_a)}], \qquad (7)$$

$$\sin^2(k d_a) = \frac{U(\omega) + (U^2(\omega) + 16\gamma_m^2 \omega_1^4)^{1/2}}{8\gamma_m^2 \omega_1^4}, \qquad (8)$$

where we have set $U(\omega) = 4\gamma_m^2 \omega_1^4 - (\omega^2 - \omega_0^2)^2 - \omega^2 \Gamma_I^2$. These relations allow to exhibit the dispersion relation $\omega(k)$ of plasmons and to calculate the transmission probability $\Im_m(\omega) = \exp[-2\alpha_m L]$ of plasmons through a chain of length $L = (N-1)d_a + d$. In order to focus our attention on the intrinsic properties of the chain we assume, in what follow, that the chain is perfectly coupled to both reservoirs. The plasmon resonance frequency $\omega_0$ of the chain embedded in a host fluid of permittivitty $\tilde{\varepsilon}$, is calculated from the Fröhlich condition $\text{Re}[\varepsilon] = -2\tilde{\varepsilon}$. In this work the host fluid we consider is ethylene glycol (EG) and its dielectric permittivity is well modeled by the lossless Debye model $\tilde{\varepsilon} = \varepsilon_\infty + \frac{\varepsilon_S - \varepsilon_\infty}{1 + \tau^2 \omega^2}$, $\tilde{\tau}$, $\varepsilon_S$ and $\varepsilon_\infty$ being the Debye relaxation time, the static dielectric constant at low frequency and the permittivity at high frequencies respectively. As for the plasma frequency, it is given by the usual relation $\omega_p = (\rho_e e^2 / \varepsilon_0 m^*)^{1/2}$ where $\rho_e$, $e$, $\varepsilon_0$ and $m^*$ denote the electronic density, the charge of electrons, the permittivity of vacuum and the effective mass of electrons. The coupling strength between adjacent dipoles is obtained from the usual relation $\omega_1^2 = \rho_e e^2 / (4\pi m^* \varepsilon_0 \tilde{\varepsilon} d_a^3)$.

In Fig.2 is plotted the dispersion relation of a chain of 10 nm particles consisting of Cu nanoparticles 10 nm diameter dispersed at 0,6% vol. in EG (i.e. with an average separation distance of 35 nm). This configuration corresponds to the type of nanofluid studied by Eastmann et al. [1] which have revealed ,without additive product, an enhancement of the EG thermal conductivity of 14% at this concentration. The physical properties of copper used to calculate the transport properties were $\rho_e = 8,5 \times 10^{28} m^{-3}$ [16], $\gamma = 1,38 \times 10^{13} s^{-1}$ [17], $m^* = 1,42 \times m_e$ [18] ($m_e$ being the mass of free electrons) and parameters used to model the dielectric propertied of EG [19] were $\varepsilon_S = 40$, $\varepsilon_\infty = 37$ and $\tilde{\tau} = 0,38 \times 10^{-9} s$.

The results for the transmission probability of modes are shown in Fig. 3. There is a relatively sharp frequency window for the energy transfer through the chain. In particular, the bandwidth for the transmission of L-modes is twice as large as that of the T modes. Also the magnitude of the transmission probability for the longitudinal modes is about three times more important than that of transversal modes. Hence L modes are the main contributors for the heat transfer through a nanoparticle chain. Results show also that the transmission coefficients decrease very rapidly with the nanoparticle concentration. At 2% vol. (i.e. $d_a = 2d$), for instance, the transmission probability is one order of magnitude smaller than that we have in contact. In Fig. 4, the thermal conductance of Cu chains embedded in EG are reported as a function of the interparticle distance for several particle sizes. We see that the thermal conductance increases continuously when the separation distance between particles is decreased. However, for levels of concentration corresponding to the operating range of nanofluids (~1% vol. or $d_a \sim 3d$) the thermal conductance is of the order of $10^{-12} W.K^{-1}$. This conductance cannot explain the 10% or larger increases of the thermal conductivity of ethylene glycol observed in [1].

So far we have assumed that all nanoparticles were at rest in the host fluid. To justify this assumption and verify that our model captures well the main physical mechanims which govern near-field heat exchanges in nanofluids, we now compare the timescales of Brownian motion to that of near field interactions. The time necessary for a particle of mass m to move on a distance equal to the interparticle distance under the action of thermal fluctuations is $\tau_B = \pi \eta d_a^3 /(2 k_B T)$. On the other hand, the electromagnetic energy is transported through the near-field interactions along a chain of particles on the same distance at the group velocity $v_{g_m} = \frac{d\omega}{dk} \approx \gamma_m \omega_1^2 d / \omega_{plasmon}$ of plasmons. Hence, the time required for these evanescent waves to transport heat on a distance $d_a$ is $\tau_{NF} \approx d_a \omega_{plasmon} /(\omega_1^2 d)$. For 10 nm diameter copper particles spaced by 35 nm in ethylene glycol ($\eta = 0.159 \, kg.m^{-1}.s^{-1}$) we find $\tau_B \approx 10^{-7} s$ while $\tau_{NF} \approx 10^{-13} s$ so that the dynamic of particles can be neglected in the near-field heat transfer. A direct inspection of two timescales expressions show that the "static nanoparticles" assumption is applicable as long as the plasmons wavelength remains in the visible or in the near infrared range of electromagnetic spectrum.

In conclusion we have studied the contribution of near-field interactions to the transport of heat in weakly concentrated nanofluids when the average separation distance between particles is small in front of plasmons wavelengths. We have shown for ethylene glycol-based nanofluids containing copper nanoparticles that the surface plasmons play a minor role in the heat transfers enhancement observed in experiments. This result is obviously applicable to others nanofluids comprising metal, metal oxide and polar nanoparticles which all support either plasmons or phonons polaritons. Our results suggest that the large thermal conductivity increases in nanofluids comes from other collective effects such as those due to the self-ordered motions of particles (i.e. collective motions consisting of phonons due to the random motion of particles which can be decomposed as harmonic-oscillator-like modes) in the host fluid or due to an increase of the phonon density in presence of clusters.

**Figure captions**

Fig. 1 : Regularly spaced nanoparticles chain connected to two reservoirs kept at neighborhood temperatures. The separation distance $d_a$ between particles is smaller than the wavelength of plasmons modes supported by the chain.

Fig. 2 : Dispersion relation for the longitudinal (L) and transverse (T) collective plasmon modes for a chain of 10 nm diameter copper particles spaced by 35 nm (0,6% vol.) in ethylene glycol.

Fig. 3 : Transmission probability of longitudinal (a) and transversal (b) plasmons through a chain of ten copper particles 10 nm diameter in ethylene glycol for several separation distances.

Fig. 4 : Near-field thermal conductance in linear chains of 10 copper particles of different size dispersed in ethylene glycol versus the separation distance.

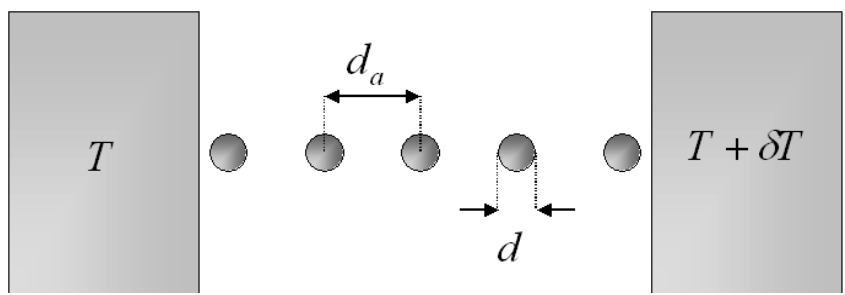

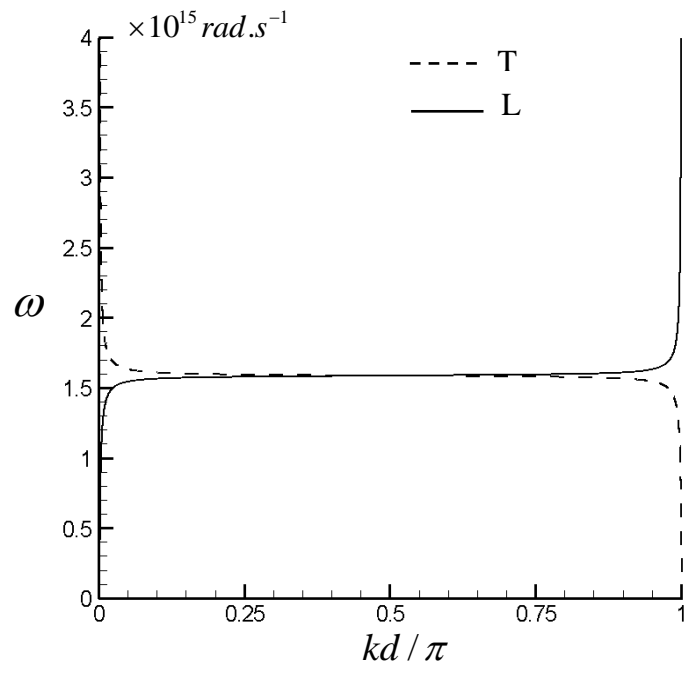

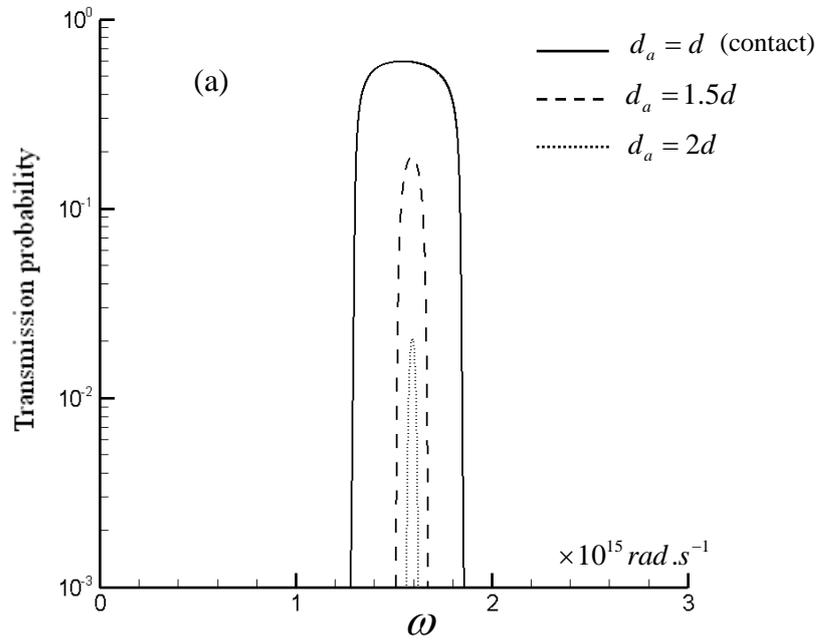

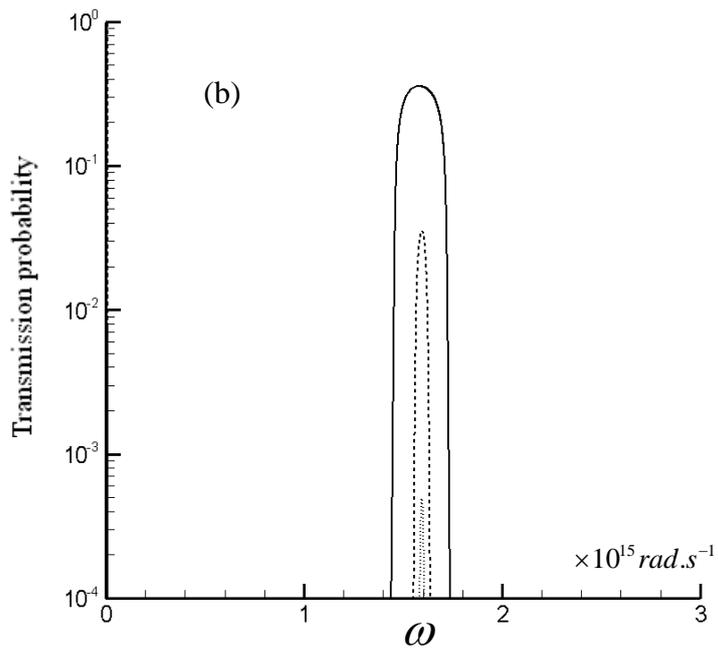

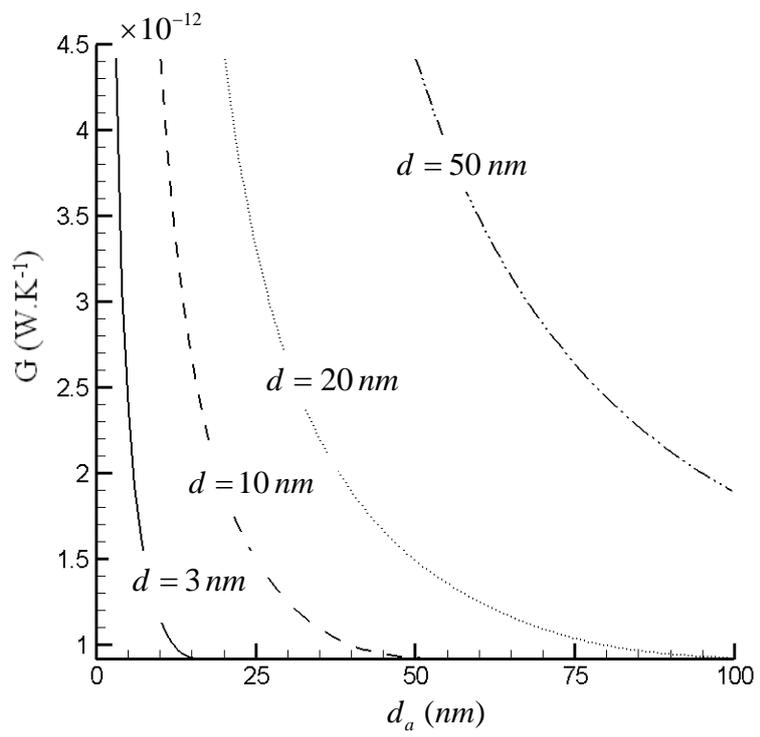